\documentclass{iopjournal}
\usepackage{xcolor}
\usepackage{amsmath}
\usepackage{latexsym}
\usepackage{amssymb}
\usepackage{amsfonts}
\usepackage{mathtools}
\usepackage{bm}
\usepackage{amsthm}
\usepackage{graphicx}
\usepackage{changes}
\usepackage[normalem]{ulem}
\usepackage{natbib}
\usepackage{appendix}
\usepackage{enumitem}
\usepackage{caption}
\usepackage{subcaption}

\def \be{\begin{equation}}
\def \bea{\begin{eqnarray}}
\def \eea{\end{eqnarray}}
\def \ee{\end{equation}}

\begin{document}
\title{Mode decomposition methods for the analysis of black hole ringdown signals}

\author{Andrzej Krolak$^1$, Massimo Tinto$^2$}

\affil{$^1$Institute of Mathematics, Polish Academy of Sciences, 00-656 Warsaw, Poland}
\\
\affil{$^2$Divis\~{a}o de Astrof\'{i}sica, Instituto Nacional de Pesquisas Espaciais, S. J. Campos, SP 12227-010, Brazil}
  
\email{krolak@impan.pl, massimo.tinto@gmail.com}
  
\date{\today}

\begin{abstract}
The ringdown phase of a binary black hole merger encodes fundamental information about the remnant through its quasinormal mode (QNM) spectrum. Extracting multiple modes from gravitational-wave data is essential for black-hole spectroscopy but remains challenging due to the short duration of the signal, limited signal-to-noise ratio (SNR), and interference between modes. In this work, we investigate the applicability of Empirical Mode Decomposition (EMD) and Variational Mode Decomposition (VMD) to the analysis of post-merger gravitational-wave signals. Using Monte Carlo simulations of noisy ringdown signals composed of pairs of QNMs, we assessed the ability of these methods to separate the two modes and estimate their frequencies. The instantaneous frequency is obtained via the Hilbert transform (HT) and our new proposed method, named instantaneous frequency and amplitude determination (IFAD). We analyze performance over a wide range of SNR values relevant to current and future gravitational-wave detectors. Our results show that both EMD and VMD can resolve multiple modes, and VMD provides significantly more accurate frequency estimates. We also introduce a modified instantaneous frequency estimator that improves accuracy over the Hilbert transform. The study quantifies the conditions under which two-mode resolution is feasible and highlights the limitations imposed by closely spaced mode frequencies and short signal duration. These results are relevant for current observations and for future high-SNR detections expected from space-based and next-generation ground detectors.
\end{abstract}

\section{Introduction}
\label{SecI}

The detection of gravitational waves from binary black hole mergers has opened a direct observational window onto the strong-field dynamics of spacetime. Following the merger, the remnant black hole relaxes to a stationary Kerr configuration through the emission of gravitational radiation known as the \emph{ringdown}. In general relativity, this phase is described as a superposition of exponentially damped sinusoids---the quasinormal modes (QNMs)---whose frequencies and damping times depend solely on the mass and spin of the remnant~\citep{Berti2025Spectroscopy,Destounis_2024}. The measurement of multiple QNMs underpins black hole spectroscopy, providing a direct test of the Kerr nature of astrophysical black holes. Resolving at least two independent modes enables internal consistency tests, as all mode frequencies must correspond to the same underlying black hole parameters. This makes ringdown observations a powerful probe of strong-field gravity and potential deviations from general relativity~\citep{Chung_2024,chung2025probingquadraticgravityblackhole}.

However, in practice, resolving multiple modes remains a challenge. Ringdown signals are short-lived and often have moderate signal-to-noise ratios (SNRs), particularly for current-generation detectors. Subdominant modes are typically weakly excited and may lie close in frequency to the dominant mode, hindering their separation. In addition, early-time data are affected by nonlinear merger dynamics, while late-time data are noise dominated, leading to significant uncertainties in frequency estimation~\citep{siegel2023ringdowngw190521hintsmultiple}. Even in recent high-SNR observations from the fourth observing run (O4), robust identification of multiple modes remains difficult~\citep{kw5g-d732,6c61-fm1n,theligoscientificcollaboration2026gwtc40testsgeneralrelativity}. These limitations motivate the development of robust, minimally model-dependent techniques for extracting mode content from noisy ringdown signals.

Traditional approaches based on matched filtering or parametric fits rely on accurate waveform models and can be sensitive to modeling assumptions. By contrast, adaptive signal decomposition methods offer a flexible framework for analyzing nonstationary and nonlinear signals such as gravitational-wave ringdowns. In this work, we investigate two such methods: Empirical Mode Decomposition (EMD) and Variational Mode Decomposition (VMD). Both approaches decompose a signal into a small number of intrinsic mode functions, each associated with a narrowband oscillatory component that can be interpreted as an approximation to an individual QNM. To extract frequency information, we employ the Hilbert transform (HT)
and a new method, named instantaneous frequency and amplitude determination (IFAD),
which provide time-resolved estimates of instantaneous frequency.

A central challenge in this framework is the reliable estimation of instantaneous frequency. Hilbert-based estimates are known to be unstable at early and late times due to boundary effects and noise contamination. Moreover, the rapid damping of ringdown signals limits the number of observable oscillation cycles, further complicating the extraction of robust frequency estimates. These issues are particularly severe for closely spaced modes or modes with significantly different amplitudes.
To address these challenges, we systematically assess the performance of EMD and VMD combined with HT and IFAD to extract QNM frequencies from noisy signals. We analyze synthetic ringdown waveforms consisting of pairs of modes embedded in Gaussian noise across a broad range of SNRs relevant to current and future detectors. This framework enables a quantitative study of the conditions under which two-mode resolution is achievable, as well as a comparison of decomposition and frequency estimation strategies. Our results provide insight into the applicability of adaptive decomposition methods to black-hole spectroscopy, particularly in the high-SNR regime expected for next-generation gravitational-wave detectors.

The paper is organized as follows. In Section~\ref{sec:methods}, we introduce the Empirical and Variational Mode Decomposition methods (Sections~\ref{ssec:EMD} and~\ref{ssec:VMD}). Section~\ref{ssec:IFi} presents our proposed instantaneous frequency estimator IFAD. In Section~\ref{sec:res}, we develop a Fisher-matrix-based geometric criterion for signal resolvability. Section~\ref{sec:Kerr} presents numerical results based on simulated ringdown signals in Gaussian noise.

\section{Mode decomposition methods}
\label{sec:methods}

In the last three decades, substantial efforts have been devoted to the development of data processing techniques capable of handling non-stationary time series generated by nonlinear physical phenomena. This resulted in significant improvements over traditional techniques based on Fourier and wavelet transforms. The representation of non-stationary signals in both time and frequency domains has been adopted in various fields such as speech recognition, biomedical data processing, telecommunication engineering, mechanical engineering, and seismic signal processing to name a few. In this regard, Empirical Mode Decomposition (EMD)~\citep{EMD}, Variational Mode Decomposition (VMD)~\citep{VMD}, and other techniques capable of decomposing an arbitrary time series into a finite sum of modes and a trend~\citep{IF1,IF2,FDM} provide a fundamentally new approach to time series analysis.

Their main feature is to generate an adaptive time-frequency decomposition that does not impose a fixed basis set on the data. As a result, these new techniques are not limited by the time-frequency uncertainty principle of Fourier-wavelet analysis and result in useful tools for investigating transients and nonlinear features. Since General Relativity is an inherently nonlinear theory, they promise to be powerful new tools in the analysis of GW data generated by ground- and future space-based GW detectors.

In this section, we provide a brief description of the EMD and VMD techniques, which we adopted to analyze time series containing monochromatic signals whose amplitudes decay exponentially. We found that the VMD technique provides the most consistent and accurate results. In what follows, we present a description of the EMD and VMD methods and point out their differences.

\subsection{Empirical Mode Decomposition}
\label{ssec:EMD}

Empirical Mode Decomposition decomposes any given data into a finite set of narrowband intrinsic mode functions (IMFs) and a trend, both derived directly from the data. The IMFs are characterized by an instantaneous frequency that is well behaved, i.e., it is positive definite in its domain of definition, and it can be represented mathematically in the following form:
\begin{equation}
    {\rm IMF}(t) \equiv A(t) \cos(\phi(t)) \ ,
    \label{eq1}
\end{equation}
where the frequency bands of the amplitude $A(t)$ and the phase $\phi(t)$ do not overlap. The main idea behind the EMD technique~\citep{EMD} is an iterative sifting process that decomposes a given signal into a set of IMFs as defined earlier. In particular, each IMF satisfies the following requirements:
\begin{enumerate}[label=\alph*)] 
   \item The total number of its extrema and the number of zero crossings are equal or differ at most by one;
   \item At any data location $s(t)$, the mean value between the envelope identified by the local maxima $IMF_{\max}(t)$ and the envelope defined by the local minima $IMF_{\min}(t)$ is zero.
\end{enumerate}

From these two properties of an IMF, the sifting process to identify an IMF from a given signal $s(t)$ proceeds as follows:
\begin{enumerate}
   \item Identify all local extrema and connect all maxima and minima with a cubic spline to form the upper and lower envelopes;
   \item The mean of the two envelopes is then subtracted from the data to obtain their difference: 
   \begin{equation}
    h(t) \equiv s(t) - \frac{s_{\max}(t) + s_{\min}(t)}{2} \ .
    \label{1stIteration}
   \end{equation}
    \item Consider $h(t)$ as the data and repeat steps (1) and (2) iteratively until the mean of the envelopes becomes sufficiently close to zero. The final $h(t)$, which we now call $y_1(t)$, is the first IMF since it satisfies the two criteria, (a) and (b), defined earlier for an intrinsic mode function.
\end{enumerate}
The residue, $r_1(t) \equiv s(t) - y_1(t)$, is then treated as new data and subjected to the sifting process described above, which produces the second IMF from $r_1(t)$. The procedure continues until either the recovered IMF or the residual data is small enough. Once all of the oscillating IMFs are subtracted from the data, the final residual component is either equal to zero (or very close to it) or is a non-oscillatory term representing the trend of the data. At the end of this iterative process, the signal $s(t)$ can be expressed in the following form:
\begin{equation}
    s(t) = \sum_{i=1}^{N} y_i(t) + r_N(t) \ ,
    \label{Decomp}
\end{equation}
where $N$ is the total number of IMFs and $r_N(t)$ is the trend of the signal. The advantage of this representation of a time series is that each IMF contains an instantaneous amplitude and frequency, which can be extracted using the Hilbert transform or other accurate techniques (such as~\cite{IF1,IF2}).

Although the EMD decomposition technique has been successfully applied to many data sets, it suffers from some issues. To start with, the endpoints of a time series are in general not extrema, making the identification of the upper and lower envelopes difficult. In addition, the use of cubic spline interpolation to construct the envelopes may introduce additional oscillations that are not present in the data. This in turn will result in IMFs that do not accurately reflect the nature of the data they are supposed to represent.

\subsection{Variational Mode Decomposition}
\label{ssec:VMD}

Variational Mode Decomposition is a non-recursive adaptive signal decomposition method that represents a given signal as a superposition of a predefined number of band-limited intrinsic mode functions (IMFs), called modes~\citep{VMD}. In contrast to EMD, VMD is formulated as a variational optimization problem, which ensures mathematical rigor and alleviates issues such as mode mixing and sensitivity to noise.

Let $s(t)$ be a real-valued signal. The objective of VMD is to decompose $s(t)$ into $N$ modes $\{y_k(t)\}_{k=1}^N$, each associated with a center frequency $\omega_k$. Each mode is assumed to be narrowband and compact around its center frequency in the spectral domain. To quantify the bandwidth of a mode, its analytic signal is constructed using the Hilbert transform, and its frequency content is shifted to the baseband by modulation with $e^{-j\omega_k t}$. The bandwidth is then estimated by the norm $L_2$ of the temporal gradient of the demodulated signal. The decomposition is obtained by solving the following constrained variational problem:
\begin{equation}
\min_{\{y_i\},\,\{\omega_i\}} \quad 
\sum_{i=1}^N
\left\|
\frac{dy^a_i(t)}{dt}
e^{-j \omega_i t}
\right\|_2^2, \quad
\text{subject to} \quad 
\sum_{i=1}^N y_i(t) = s(t),
\label{eq:vmd}
\end{equation}
where $y^a_i(t)$ is the analytic signal of $y_i(t)$.

To solve the constrained optimization problem in \eqref{eq:vmd}, an augmented Lagrangian is introduced:
\begin{equation}
\begin{aligned}
\mathcal{L}(\{y_i\}, \{\omega_i\}, \lambda) =
& \, \alpha \sum_{i=1}^N
\left\|
\frac{dy^a_i(t)}{dt}
e^{-j \omega_i t}
\right\|_2^2 \\
& + \left\| s(t) - \sum_{i=1}^N y_i(t) \right\|_2^2
+ \left\langle \lambda(t), s(t) - \sum_{i=1}^N y_i(t) \right\rangle,
\end{aligned}
\end{equation}
where $\lambda(t)$ is the Lagrange multiplier enforcing the reconstruction constraint and $\alpha$ is a quadratic penalty parameter controlling the bandwidth of the modes. The inner product is defined as $\left\langle p(t), q(t) \right\rangle = \int^{\infty}_{-\infty} p^*(t) q(t) \, dt$ and $\left\|p(t)\right\|_2^2 = \left\langle p(t), p(t) \right\rangle$.

The augmented Lagrangian is minimized using the alternating direction method of multipliers \citep{Hestenes1969}. In each iteration, the modes $y_k$ and their center frequencies $\omega_k$ are updated alternately in the Fourier domain, while the Lagrange multiplier is updated via dual ascent. The update of each mode corresponds to a Wiener filtering operation, and the center frequencies are estimated as the weighted average of the power spectrum of the corresponding mode. The iterative process continues until a convergence criterion is met, typically based on the relative change of the modes between successive iterations. Due to its variational formulation, the convergence of VMD is well controlled and largely independent of the signal length.

Because of its robustness to noise, clear spectral separation, and solid theoretical foundation, VMD has been successfully applied in a wide range of signal processing applications, including denoising, time--frequency analysis, biomedical signal processing, and mechanical fault diagnosis.

\subsection{Instantaneous Frequency and Amplitude Determination}
\label{ssec:IFi}

Once the intrinsic mode functions (IMFs) defining a given time series have been extracted, the question of how to determine their instantaneous frequency and amplitude arises. The standard approach in the literature to derive these quantities is to use the Hilbert Transform (HT). Unfortunately, this method has several drawbacks. The major one is related to the finite duration of the time series. Because the HT is a convolution that requires infinite data, applying it to finite-length signals causes large distortions and ``ringing'' at the beginning and end of the record (edge effects), penalizing the accuracy of the determination of the physical quantities of interest.

Here, we present an alternative technique that avoids the issues associated with the use of the HT and estimates the two quantities that characterize an IMF with high precision. Since an IMF, $y(t)$, can be represented in the following mathematical form:
\begin{equation}
    y(t) \equiv A(t) \cos(\phi(t)) \ , \ t \in [0, T]
    \label{mode}
\end{equation}
the procedure for obtaining its amplitude and instantaneous frequency at time $t$, $f(t) = \frac{1}{2\pi}\frac{d\phi(t)}{dt}$, is based on the use of interpolation while following the procedure highlighted below.

First, we identify the location of the maxima of $|y(t)|$. We then interpolate these points, retaining only the samples between the first and last extrema located in the defined instances $t_0$ and $t_n$, respectively. If we now denote by $\hat A(t)$ our estimate of the interpolated amplitude in the interval $[t_0, t_n]$, we can obtain $\cos(\phi(t))$ in this interval using the following expression:
\begin{equation}
    C(t) \equiv y(t)/{\hat A(t)} \ .
\end{equation}
Although it has already been shown~\citep{CEMD,FreyOsorio} that $C(t)$ can then be used to determine the instantaneous phase, $\phi(t)$, and the frequency, $f(t)$, of the IMF mode without using HT, here instead we will proceed by using the standard definition of frequency as the inverse of the period. In fact, this can be obtained by relying on the time locations of the extrema and the zero-crossings of $C(t)$. This approach is physically well defined and more precise than procedures based on the Hilbert transform or other techniques.

After ordering the time-sequence locations of the extrema and the zero-crossings, one can compute the sequence of periods from them. By taking the inverse of this sequence, we obtain the corresponding instantaneous frequency of the IMF considered. As the instantaneous frequency is estimated at a discrete number of sampled points, we can interpolate them to obtain the instantaneous frequency at any time within the time interval of interest.

As an example application of our proposed method for estimating the instantaneous amplitude and frequency of a mode, which we here christen as the \emph{Instantaneous Frequency and Amplitude Determination} (IFAD) technique, we consider three possible waveforms:
\begin{align}
y_{sd}(t) &= A_{sd}(t)\cos(\phi_{sd}(t)), \quad \text{where: } A_{sd}(t) = e^{-t / \tau}, \ \phi_{sd}(t) = 2\pi f_o t + \pi \dot{f}_o t^2 + \phi_o, \\
y_{qnm}(t) &= A_{qnm}(t)\cos(\phi_{qnm}(t)), \quad \text{where: } A_{qnm}(t) = e^{-t / \tau}, \ \phi_{qnm}(t) = 2\pi f_o t + \phi_o, \\
y_{\text{chirp}}(t) &= A_{\text{chirp}}(t)\cos(\phi_{\text{chirp}}(t)), \quad \text{where: } \nonumber \\
A_{\text{chirp}}(t) &= \mathcal{M}\left(\frac{t_c - t}{5\mathcal{M}}\right)^{-1/4}, \ \phi_{\text{chirp}}(t) = \phi_o - 2 \left(\frac{t_c - t}{5\mathcal{M}}\right)^{5/8},
\end{align}
where $y_{sd}$ is a sinusoidal damped signal with damping time $\tau$, frequency $f$ and frequency drift $\dot{f}$; $y_{qnm}$ is a Kerr black hole quasinormal mode; and $y_{\text{chirp}}$ is a chirping gravitational wave signal emitted by a binary system.

For all signals, $\phi_o$ is a constant phase. In the case of the quasinormal mode signal, the frequency and the damping time are determined by the mass and spin of the Kerr black hole. In the case of the chirping signal, $\mathcal{M}$ is the chirp mass defined as:
\begin{equation}
\mathcal{M} = \frac{(m_1 m_2)^{3/5}}{(m_1 + m_2)^{1/5}},
\end{equation}
where $m_1$ and $m_2$ are the masses of the binary system, while $t_c$ and $k_o$ are constants. The instantaneous frequencies $f(t)$ of the three signals are given by:
\begin{align}
f_{sd}(t) &= f_o + \dot{f}_o t, \\
f_{qnm}(t) &= f_o, \\
f_{\text{chirp}}(t) &= \frac{1}{8 \pi} \frac{1}{\mathcal{M}} \left(\frac{t_c - t}{5\mathcal{M}}\right)^{-3/8}.
\end{align}

\begin{figure}[htbp]
    \centering
    \begin{subfigure}[b]{0.32\textwidth}
        \centering
        \includegraphics[width=\textwidth]{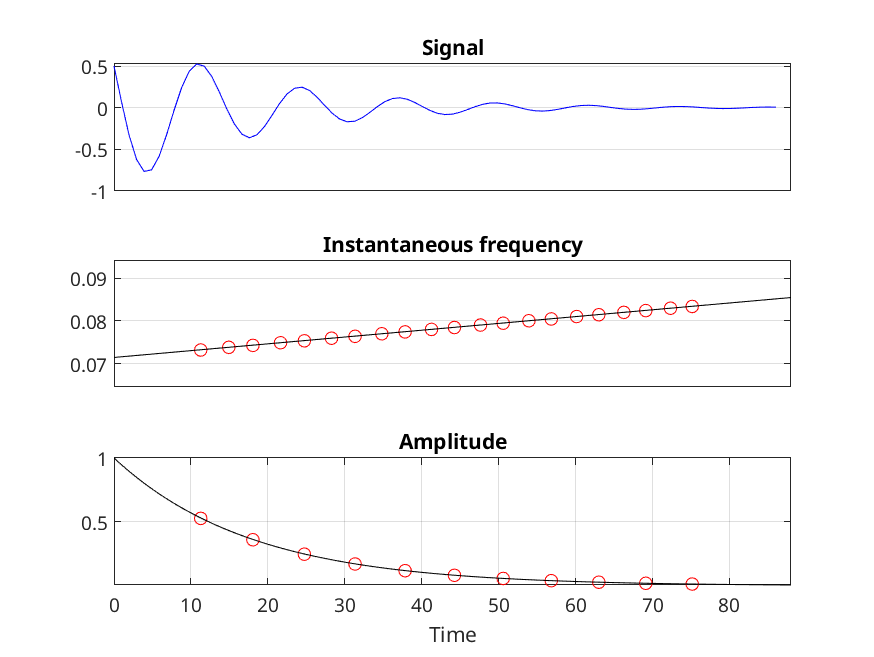}
        \caption{Damped sinusoid signal with frequency drift.}
    \end{subfigure}
    \hfill
    \begin{subfigure}[b]{0.32\textwidth}
        \centering
        \includegraphics[width=\textwidth]{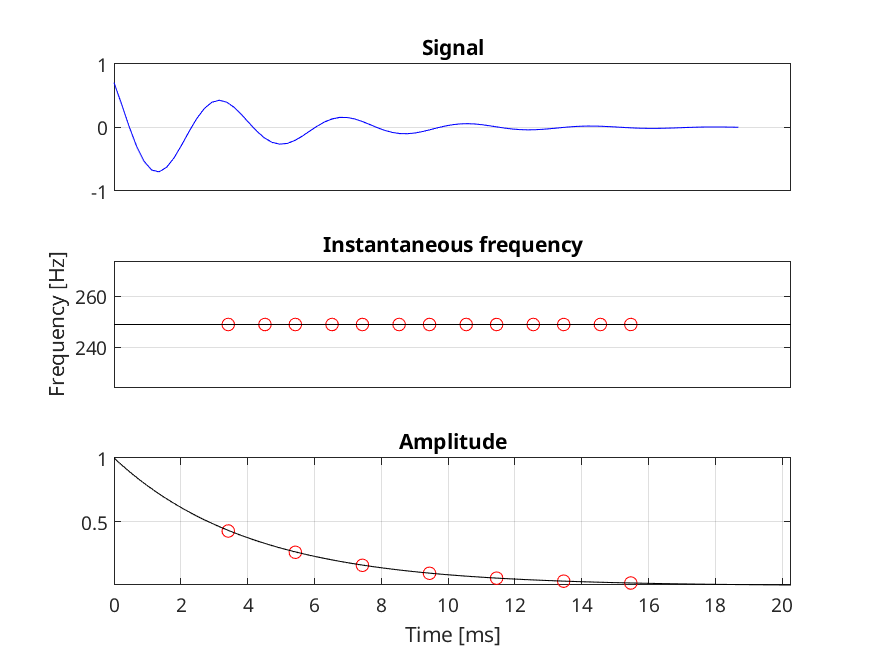}
        \caption{Quasinormal mode of Kerr black hole.}
    \end{subfigure}
    \hfill
    \begin{subfigure}[b]{0.32\textwidth}
        \centering
        \includegraphics[width=\textwidth]{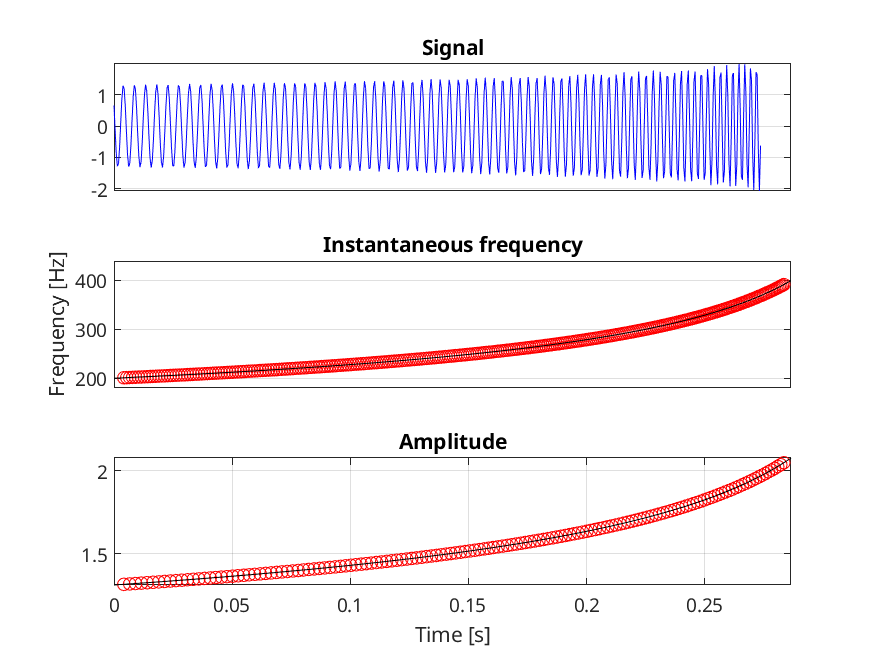}
        \caption{Chirping gravitational wave signal from a binary system.}
    \end{subfigure}   
    \caption{Top panels represent the signal analyzed. In the middle panel, we plot the true frequency of the signal as a continuous line alongside our estimated values represented by the 'o' symbols. In the bottom panel, we plot the true amplitude (continuous line) alongside our estimated values ('o' symbols).}
    \label{fig:InstFandAmp}
\end{figure}

In Figure~\ref{fig:InstFandAmp}, we present the results of the application of our method to the three signals given above.

For the case of the quasinormal mode, we analyzed the fundamental mode and assumed mass $M = 68.343\ \mathrm{M}_{\odot}$ and spin $a = 0.68$. For the case of the chirping signal, we assumed that the masses of the two components of the binary system are equal to each other and to $1.4\ \mathrm{M}_{\odot}$. The starting frequency $f_i = f_{\text{chirp}}(0)$ at time $t = 0$ is taken to be $200$~Hz. Our instantaneous frequency and amplitude estimation methods achieve very high accuracy, as can be visually inferred since both the continuous lines and the 'o' symbols lie perfectly on top of each other. This performance is expected because the accuracy that can be achieved by this method is primarily determined by the numerical precision in locating the extrema and zero crossings.

\section{Resolution of damped sinusoid signals and quasinormal modes}
\label{sec:res}

Consider the following signal $s(t)$ that consists of the sum of two damped sinusoids:
\begin{equation}
\label{eq:sig2sin}
s(t) = h_1 e^{-t / \tau_1} \cos(2\pi f_1 t +\phi_1) + h_2 e^{-t / \tau_2} \cos(2\pi f_2 t + \phi_2), \quad t = 0,\dots, N-1 \ ,
\end{equation}
where $N$ is the number of data points, $h_1, h_2$ and $\phi_1, \phi_2$ are the amplitudes and phases of the two components, respectively. The parameters $\tau_1, \tau_2$ and $f_1, f_2$ are the damping times and frequencies of the two components, respectively. We call the two components of the signal $s(t)$ \emph{modes}.

Section IVC1 of~\cite{isi2021} contains an excellent discussion of the problem of resolving the two modes of the signal \eqref{eq:sig2sin} in noise. In particular, it underlined that this is a problem of resolution in the two-dimensional space spanned by frequency $f$ and damping time $\tau$. For example, one can resolve damped sinusoids with an arbitrarily small frequency difference, provided that the damping times are sufficiently separated. The discussion in~\cite{isi2021} was in the context of Bayesian analysis, and the following criterion for the resolvability of damped sinusoids (\cite{isi2021}, Eq.~(66)) was proposed under the assumption that the posteriors of the parameters do not show significant correlations among the frequencies $f$ and damping times $\tau$:
\begin{equation}
\label{eq:recritIW}
r^2 \equiv \frac{(f_1 - f_2)^2}{\sigma^2_{f_1} + \sigma^2_{f_2}} + \frac{(\tau_1 - \tau_2)^2}{\sigma^2_{\tau_1} + \sigma^2_{\tau_2}} \geq 1,
\end{equation}
where $\sigma_\theta$ is the standard deviation of the posterior samples for the parameter $\theta$. The criterion can also be used in the context of maximum likelihood estimation where $\sigma_\theta$ are the standard deviations of the parameters $\theta$ obtained from the Fisher matrix.

Here, in the context of maximum likelihood estimation, we propose the following geometric resolvability criterion in terms of the Fisher matrix, valid when the parameters $f$ and $\tau$ can be arbitrarily correlated:
\begin{equation}
\label{eq:recrit}
\rho^2 \equiv d\theta \cdot \bar{\Gamma} \cdot d\theta^T \geq 1,
\end{equation}
where $d\theta = [\Delta f, \Delta \tau]$, $\Delta f = f_1 - f_2$, $\Delta\tau = \tau_1 - \tau_2$, and $\bar{\Gamma}$ is the Fisher matrix projected into the two-dimensional parameter space spanned by the parameters $d\theta$\footnote[1]{We obtain the $2 \times 2$ matrix $\bar{\Gamma}$ by the following procedure. First, we calculate the full $8 \times 8$ Fisher matrix $\Gamma$ for the signal $s(t)$ for a given signal-to-noise ratio. Then we transform the matrix $\Gamma$ to the parameters $\theta = [h_1, \phi_1, f_1, \tau_1, h_2, \phi_2, \Delta f, \Delta\tau]$. We take the inverse of the resulting matrix, which gives the $8 \times 8$ covariance matrix $C$. We take the $2 \times 2$ sub-matrix $\bar{C}$ of $C$ corresponding to the parameters $\Delta f$ and $\Delta\tau$. The inverse of $\bar{C}$ is our desired matrix $\bar{\Gamma}$.}.

\begin{figure}[ht]
\centering
\includegraphics[width=0.6\textwidth]{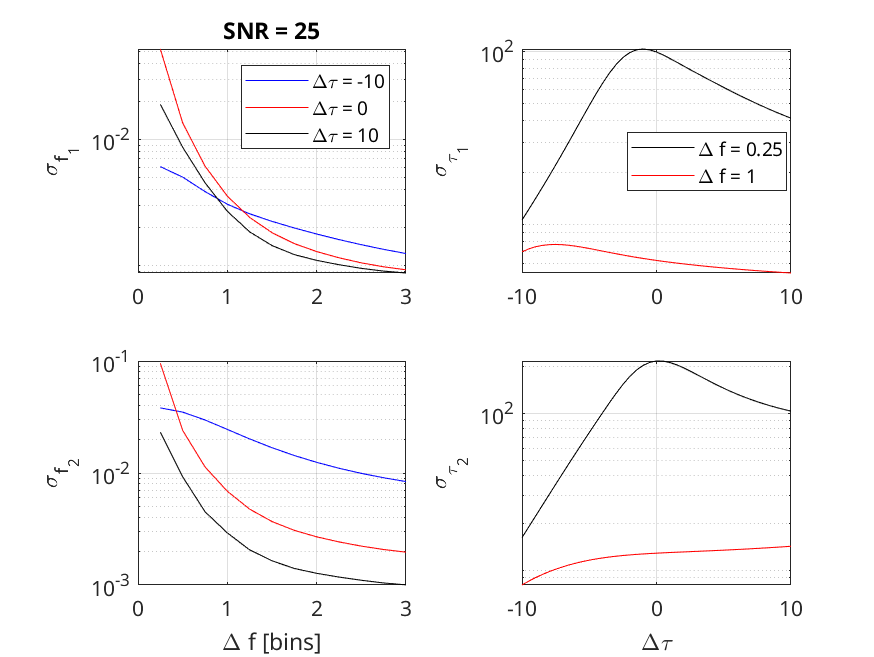}
\caption{Standard deviations of the damping times and frequencies of the signal \eqref{eq:sig2sin}, estimated using the Fisher matrix formalism. The left panels show the standard deviations of the frequencies as a function of the frequency separation between the two components. Results are presented for three different values of $\Delta\tau$, corresponding to differences in the damping times of the components. The right panels show the corresponding standard deviations of the damping times as a function of the damping-time separation, for two different values of $\Delta f$, representing the frequency differences between the components. The bin width in frequency, $df$, is given by $1/N$, where $N$ denotes the number of data points in the signal.}
\label{fig:res_fisher}
\end{figure}

We study the criterion \eqref{eq:recrit} for various choices of parameter separation $f$ and $\tau$ in the signal \eqref{eq:sig2sin}. First, we present the standard deviations $\sigma_f$ and $\sigma_\tau$ of the parameters $f$ and $\tau$ for the two modes of the signal for various separations between them. This is shown in Figure~\ref{fig:res_fisher}. We assume that the signal is buried in zero-mean Gaussian noise, and we set the signal-to-noise ratio (SNR) equal to 25. We take the length of the signal $s(t)$ to be 5 times the damping time $\tau_1$ of the first mode. Within this length, we retain 99.99\% of the total power of the damped sinusoid. Standard deviations are calculated from the 8-dimensional Fisher matrix for the parameters $\Theta = [h_1, \phi_1, f_1, \tau_1, h_2, \phi_2, f_2, \tau_2]$ of the signal.
We see from the plots that the standard deviations increase as the parameters of the two modes get closer. 

\begin{figure}[htbp]
    \centering
    \begin{subfigure}[b]{0.45\textwidth}
        \centering
        \includegraphics[width=\textwidth]{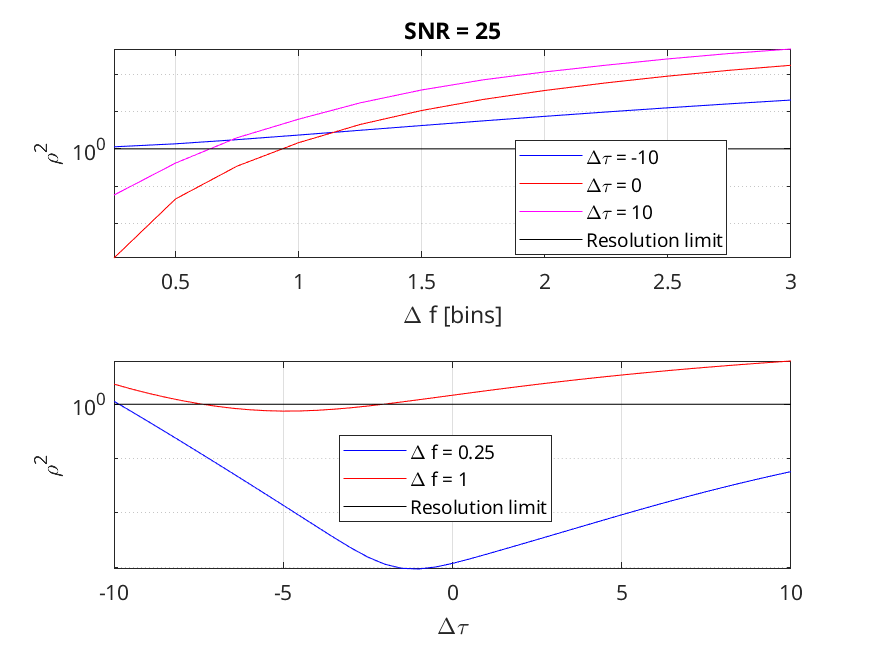}
    \end{subfigure}
    \hfill
    \begin{subfigure}[b]{0.45\textwidth}
        \centering
        \includegraphics[width=\textwidth]{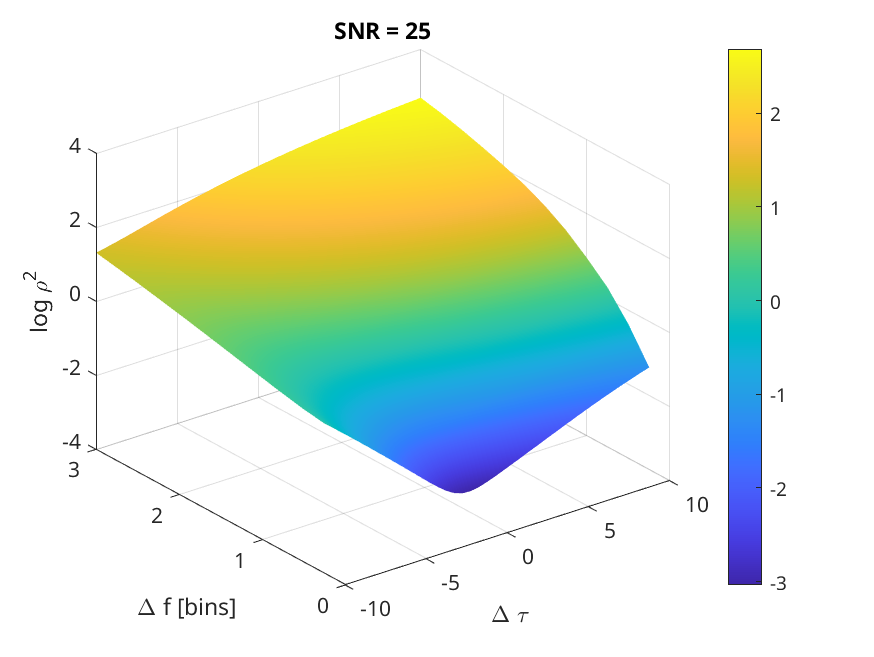}
    \end{subfigure}
    \caption{Resolvability parameter $\rho^2$ as a function of the differences of the damping times and frequencies of the two components of the signal $s(t)$. The left top panel shows $\rho^2$ as a function of the difference of frequencies of the two modes for three values of differences in damping times. The left bottom panel shows $\rho^2$ as a function of the difference of damping times of the two modes for two values of differences in frequencies. The right panel shows our resolution parameter as a function of both the frequencies and damping times differences.}
\label{fig:resolvability}
\end{figure}

In Figure~\ref{fig:resolvability}, we plot the resolution parameter $\rho^2$ (Eq.~\eqref{eq:recrit}) as a function of the separation of frequencies and damping times of the two components of the signal \eqref{eq:sig2sin}. Figure~\ref{fig:resolvability} shows that the resolution parameter $\rho^2$ increases as the separation between the parameters of the two modes increases. However, the effect is not symmetric for the parameter $\tau$. The resolution is higher for positive values of the difference $\Delta\tau$. This is because for positive $\Delta\tau$, the value of the damping parameter $\tau$ is larger for the second mode than for the first mode. The opposite holds for negative $\Delta\tau$. A larger $\tau$ yields more cycles in the signal, which directly improves resolvability.

\begin{figure}[ht]
\centering
\includegraphics[width=0.6\textwidth]{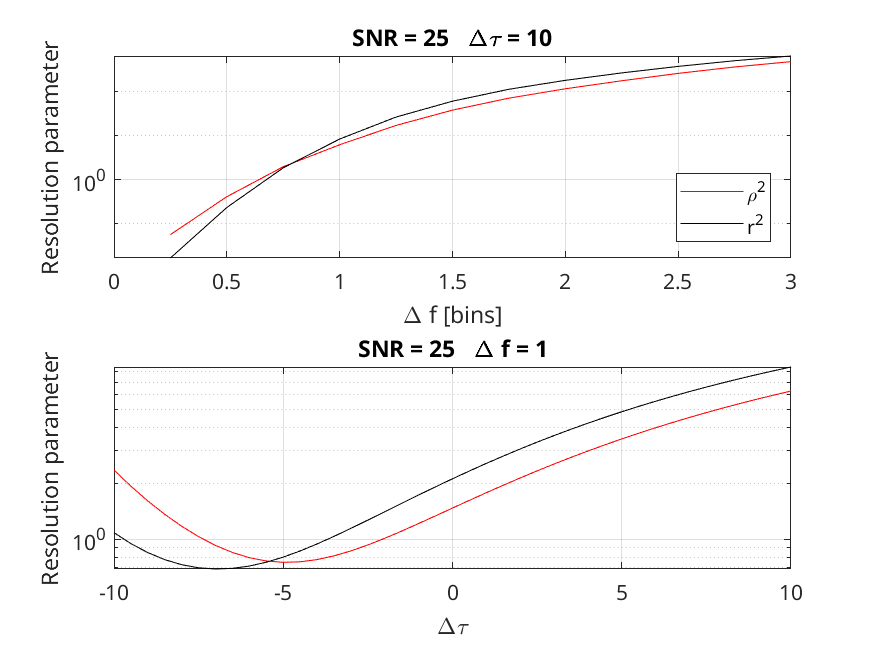}
\caption{Comparison of the resolvability criterion of Isi and Farr~\citep{isi2021} given by the parameter $r^2$ and the criterion presented in this paper given by $\rho^2$.}
\label{fig:res_compare}
\end{figure}

In Figure~\ref{fig:res_compare}, we compare the resolvability criterion of~\cite{isi2021} given by Eq.~\eqref{eq:recritIW} with the criterion given by Eq.~\eqref{eq:recrit} proposed in this paper. Figure~\ref{fig:res_compare} shows that there is qualitative agreement between the two criteria.

\begin{figure}[ht]
\centering
\begin{subfigure}{.5\textwidth}
  \centering
  \includegraphics[width=0.8\linewidth]{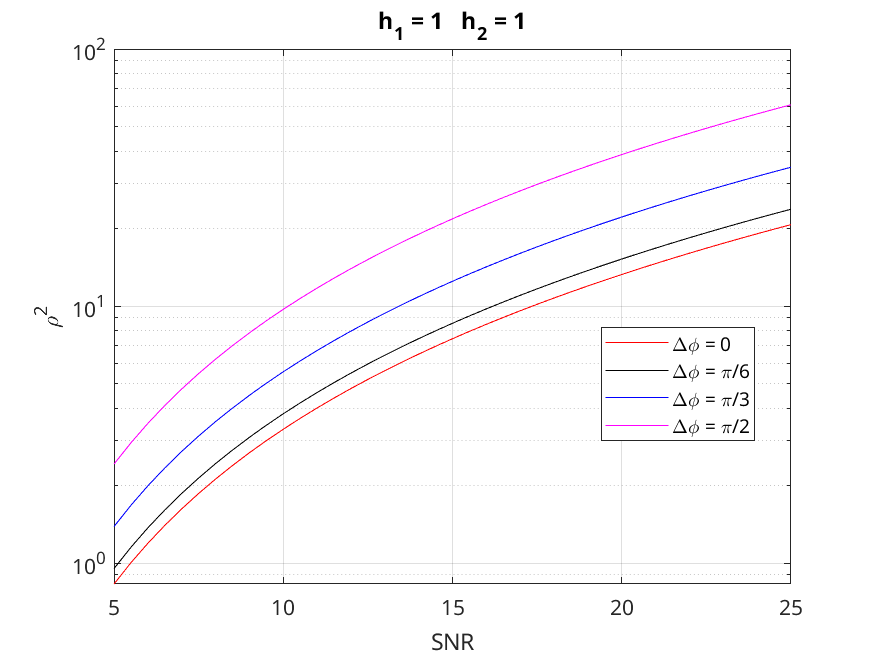}
\end{subfigure}%
\begin{subfigure}{.5\textwidth}
  \centering
  \includegraphics[width=0.8\linewidth]{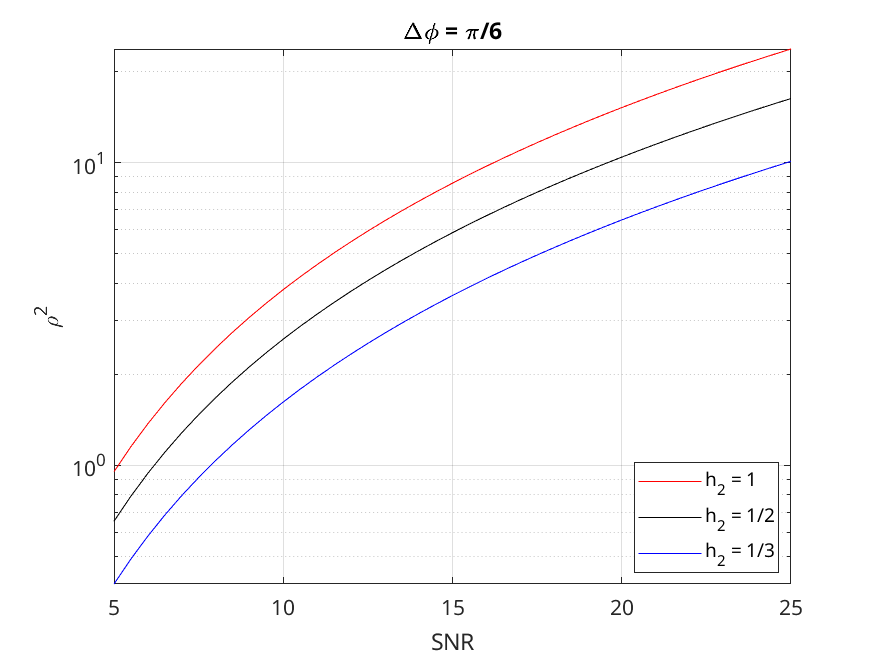}
\end{subfigure}
\caption{Resolvability parameter as a function of the signal-to-noise ratio for various amplitudes and phases of the two modes of the signal $s(t)$. The figure on the left shows the dependence of parameter $\rho^2$ on SNR depending on the relative difference in the phases of the two modes. The figure on the right shows the same dependence on the amplitude ratio of the two modes.}
\label{fig:reshophio}
\end{figure}

In Figure~\ref{fig:reshophio}, we have investigated the resolvability parameter $\rho^2$ as a function of the signal-to-noise ratio, as well as its dependence on the relative amplitudes and phases of the two components of our signal. We clearly see that resolvability increases with the signal-to-noise ratio. The resolvability also increases with an increasing difference between the phases of the two modes and with a decreasing amplitude ratio of the modes.

\subsection{Resolvability of quasinormal modes of the Kerr space-time}
\label{ssec:resKerr}

In this section, we investigate the resolvability of quasinormal modes predicted by linear perturbations of Kerr space time~\citep{PhysRevD.5.2419,PhysRevD.5.2439} using the criterion \eqref{eq:recrit}. Such modes arise as a result of the merger of black holes and are currently the subject of intensive studies~\citep{berti2025blackholespectroscopytheory}. The quasinormal modes form a countably infinite family of damped sinusoids parameterized by three parameters $\ell, m,$ and $n$, where $(\ell m)$ are angular harmonic numbers and $n$ is the overtone number. Thus, a quasinormal mode $h_{\ell mn}(t)$ has the form:
\begin{equation}
h_{\ell mn}(t) = A_{\ell mn} \exp(-t/\tau_{\ell mn}) \cos(2\pi t f_{\ell mn} + \phi_{\ell mn}),
\end{equation}
where $\tau_{\ell mn}$ and $f_{\ell mn}$ are the damping time and frequency of the mode $h_{\ell mn}(t)$, respectively. We denote any quasinormal mode $h_{\ell mn}(t)$ simply as $[\ell\ m\ n]$.

By the no-hair theorem, the damping times and frequencies of the modes are determined uniquely by two parameters of the Kerr black hole: its mass $M$ and spin $a$. The detailed relations between $\tau_k$ and $f_k$ and $M$ and $a$ obtained by perturbation theory can be found on the web page \texttt{https://pages.jh.edu/eberti2/ringdown/}. The constant amplitudes $A_{\ell mn}$ and phases $\phi_{\ell mn}$ depend in a complicated way on the initial conditions of the binary merger.

In our study, as an illustrative example, we take the parameters of the remnant black hole from the strongest gravitational-wave signal of a binary black hole (BBH) merger, the event GW250114~\citep{kw5g-d732}:
\begin{equation}
M = 68.343\,\mathrm{M}_{\odot}, \qquad a = 0.68,
\end{equation}
where $\mathrm{M}_{\odot}$ is the solar mass. We consider the resolution of the fundamental mode $[2\ 2\ 0]$, which is the strongest excited mode, and the modes $[2\ 2\ 1], [2\ 1\ 0], [2\ 0\ 0], [3\ 3\ 0], [3\ 2\ 0], [4\ 4\ 0]$, which are found to be the most excited by numerical studies of BBH mergers~\citep{PhysRevD.76.064034,PhysRevD.75.124018}.

In Table~\ref{tab:qnms}, we show the differences between the parameters $f$ and $\tau$ of the fundamental mode and the six subdominant modes listed above. The length $N$ of the signal is taken to be equal to five times the damping time of the fundamental mode. The difference $\Delta f$ in frequencies is normalized by the frequency bin $df = 1/N$, and the difference $\Delta\tau$ in damping times is normalized by the damping time $\tau_{[2\ 2\ 0]}$ of the fundamental mode. As the frequencies of the quasinormal modes are inversely proportional to the mass of the black hole and the damping times are directly proportional to it, the differences presented in Table~\ref{tab:qnms} are independent of the mass of the black hole and depend uniquely on the spin $a$.

\begin{table}[ht]
\centering
\caption{\label{tab:qnms} Differences between frequencies and damping times of the fundamental mode and six subdominant quasinormal modes of a Kerr black hole with spin $a = 0.68$. The first two columns are dimensionless differences independent of the mass of the black hole. The last two columns provide the absolute differences for a black hole with mass $M = 68.343 \, \mathrm{M}_{\odot}$.}
\vspace{5mm}
\begin{tabular}{|c|c|c|c|c|}
\hline
Mode & $\Delta f$ [bins] & $\Delta\tau$ [$\tau_{[2\ 2\ 0]}$] & $\Delta f$ [Hz] & $\Delta\tau$ [$\tau_{[2\ 2\ 0]}$ [ms]] \\
\hline
$[2\ 2\ 1]$ & 0.11 & 0.671 & 5.5 & 2.7 \\
$[2\ 1\ 0]$ & 0.70 & 0.005 & 34  & 0.02 \\
$[2\ 0\ 0]$ & 1.29 & 0.023 & 63  & 0.10 \\
$[3\ 3\ 0]$ & 2.99 & 0.026 & 146 & 0.11 \\
$[3\ 2\ 0]$ & 2.20 & 0.033 & 107 & 0.13 \\
$[4\ 4\ 0]$ & 5.89 & 0.044 & 287 & 0.18 \\
\hline
\end{tabular}
\end{table}

\begin{figure}[ht]
\centering
\includegraphics[width=0.9\textwidth]{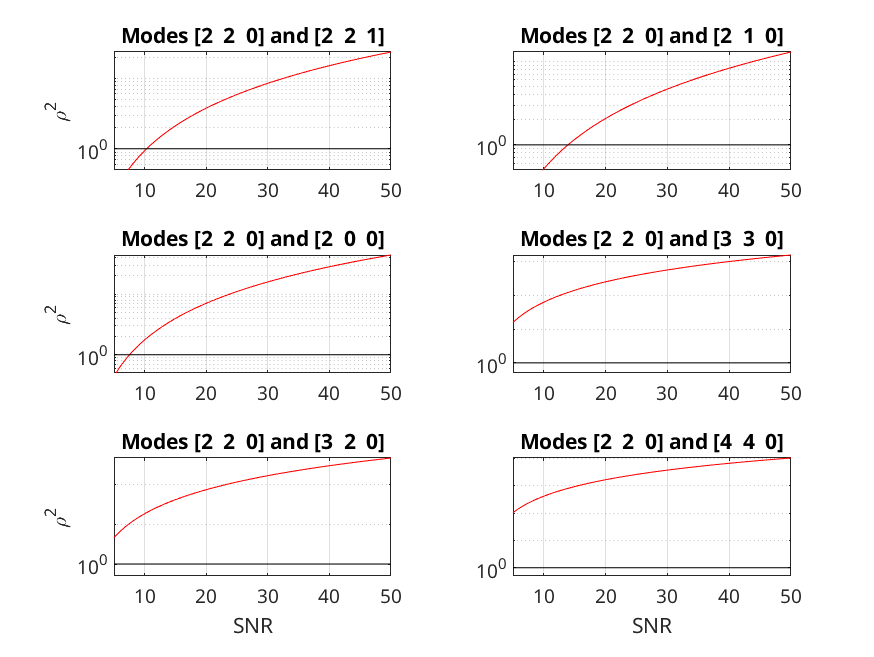}
\caption{The resolvability parameter $\rho^2$ defined by Eq.~\eqref{eq:recrit} as a function of the signal-to-noise ratio for resolving the six quasinormal modes $[2\ 2\ 1], [2\ 1\ 0], [2\ 0\ 0], [3\ 3\ 0], [3\ 2\ 0], [4\ 4\ 0]$ from the fundamental mode $[2\ 2\ 0]$. The values of $\rho^2$ are independent of the mass of the black hole and depend only on its spin $a$. The black horizontal lines denote the resolution limit of $\rho^2 = 1$.}
\label{fig:res_qnm}
\end{figure}

In Figure~\ref{fig:res_qnm}, we have calculated the parameter $\rho^2$ as a function of the signal-to-noise ratio to resolve the six modes selected from the fundamental mode for the case of the event GW250114. We see from Figure~\ref{fig:res_qnm} and Table~\ref{tab:qnms} that the dominant quantity determining resolvability is the frequency difference $\Delta f$ between the modes.

\begin{figure}[ht]
\centering
\includegraphics[width=0.5\textwidth]{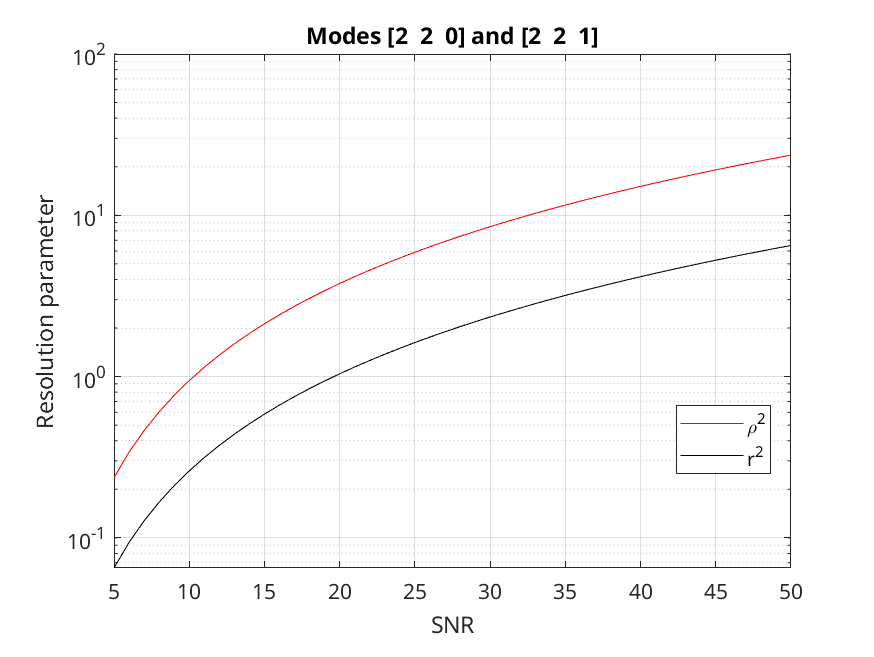}
\caption{Comparison of the resolvability criterion of~\cite{isi2021} ($r^2$) with the criterion presented in this paper ($\rho^2$) for the case of resolving the 1st overtone $[2\ 2\ 1]$ from the fundamental mode $[2\ 2\ 0]$ with the spin parameter $a$ of the remnant black hole in the event GW250114.}
\label{fig:res_compare221}
\end{figure}

In Figure~\ref{fig:res_compare221}, we compare the resolvability parameter $r^2$ (Eq.~\eqref{eq:recritIW}) with the metric $\rho^2$ (Eq.~\eqref{eq:recrit}) proposed in this paper for the specific case of modes $[2\ 2\ 0]$ and $[2\ 2\ 1]$. We see that the criterion of Isi and Farr systematically underestimates the resolvability relative to our geometric criterion in this configuration.

\section{Application of mode decomposition methods to the analysis of black hole ringdowns}
\label{sec:Kerr}

In this section, we present the results of Monte Carlo simulations aimed at estimating the frequencies of quasinormal modes of a Kerr black hole using the EMD and VMD methods. As in Section~\ref{ssec:resKerr}, we consider the resolution of the modes $[2\ 2\ 1]$, $[2\ 1\ 0]$, $[2\ 0\ 0]$, $[3\ 3\ 0]$, $[3\ 2\ 0]$, and $[4\ 4\ 0]$ in the presence of the fundamental mode $[2\ 2\ 0]$. For the analysis, we construct signals $s(t)$ consisting of pairs of modes: the fundamental mode and one of the six subdominant modes listed above. We adopt the following values for the black hole mass and dimensionless spin:
\begin{equation}
M = 68.343 \, \mathrm{M}_{\odot}, \qquad a = 0.68.
\end{equation}

For each configuration, the signal is embedded in Gaussian noise and scaled to achieve a target signal-to-noise ratio (SNR). We explore SNR values ranging from $10$ to $500$, performing $1000$ independent simulations per realization. In each run, both the EMD and the VMD methods are applied to decompose the noisy data stream into two component modes. Subsequently, the Hilbert transform~\citep{doi:10.1142/S1793536909000096} and our IFAD technique are implemented to extract the time-dependent instantaneous frequency profile of each isolated mode.

\begin{figure}[ht]
\centering
\begin{subfigure}[b]{.45\textwidth}
  \centering
  \includegraphics[width=\linewidth]{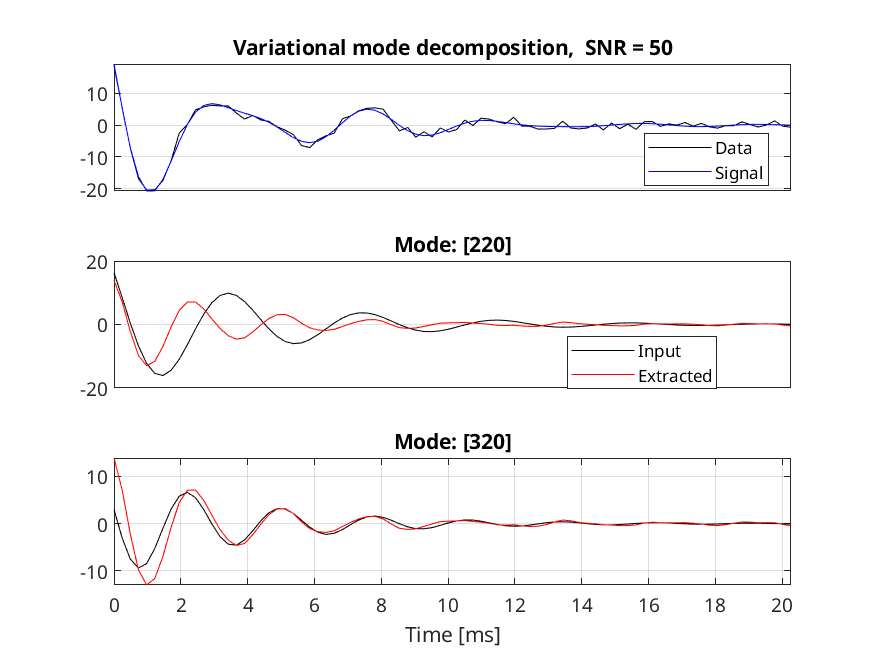}
  \caption{Top panel displays the raw input signal alongside the synthetic dataset embedded in Gaussian noise. Middle and bottom panels exhibit the isolated original modes and their corresponding variational mode decomposition reconstructions.}
  \label{sfig:modes}
\end{subfigure}
\hfill
\begin{subfigure}[b]{.45\textwidth}
  \centering
  \includegraphics[width=\linewidth]{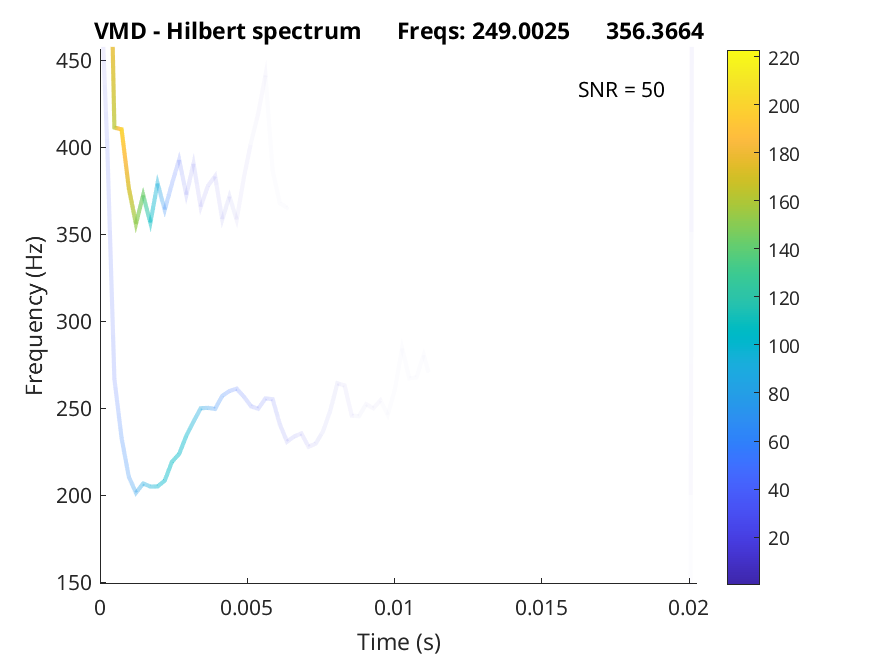}
  \caption{Hilbert spectrum of the extracted modes evaluated using the Hilbert transform. Nominal frequencies are marked at the top. The underlying trajectories are clearly distinguished.}
  \label{sfig:hilbspec}
\end{subfigure}
\caption{Application of the variational mode decomposition method to extract modes from a signal composed of two quasinormal modes in Gaussian noise at an SNR of 50.}
\label{fig:VMD_ex1}
\end{figure}

In Figure~\ref{fig:VMD_ex1}, we present an example application of the VMD technique to a signal composed of modes $[2\ 2\ 0]$ and $[3\ 2\ 0]$ added to Gaussian noise with $\mathrm{SNR} = 50$.

As illustrated in Figure~\ref{sfig:hilbspec}, the Hilbert spectrum clearly resolves the frequencies of the two constituent components. However, the raw instantaneous frequency estimate $f_{\mathrm{inst}}(t)$ is highly unstable at early times. While its precision settles during intermediate intervals, accuracy drops off dramatically as the tail end of the signal amplitude decays into the baseline noise floor. To construct robust and stabilized frequency measurements, we discard the initial $6\%$ and the final $40\%$ of the instantaneous frequency timeline. The final frequency estimate of each mode is computed as the statistical mean of the remaining intermediate window of $f_{\mathrm{inst}}(t)$.

\begin{figure}[ht]
\centering
\begin{subfigure}[b]{.5\textwidth}
  \centering
  \includegraphics[width=\linewidth]{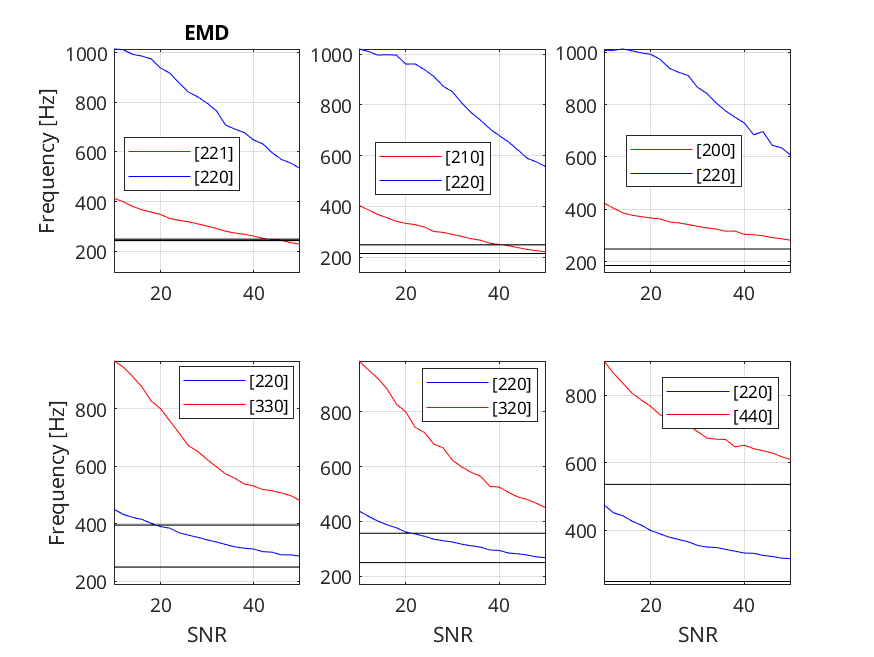}
  \caption{Low SNR regime.}
\end{subfigure}%
\hfill
\begin{subfigure}[b]{.5\textwidth}
  \centering
  \includegraphics[width=\linewidth]{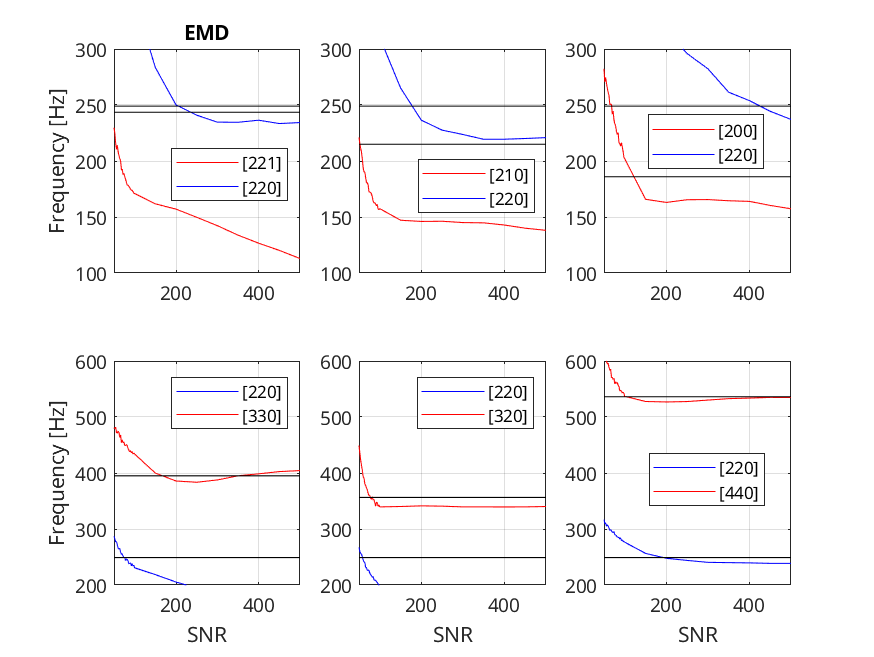}
  \caption{High SNR regime.}
\end{subfigure}
\caption{Application of empirical mode decomposition (EMD) to resolve Kerr black hole quasinormal modes in noise. Each panel shows the frequency estimate of the two modes as a function of the signal-to-noise ratio (SNR) for a signal composed of the fundamental mode and one additional mode.}
\label{fig:EMD_sim}
\end{figure}

\begin{figure}[ht]
\centering
\begin{subfigure}{.49\textwidth}
  \centering
  \includegraphics[width=\linewidth]{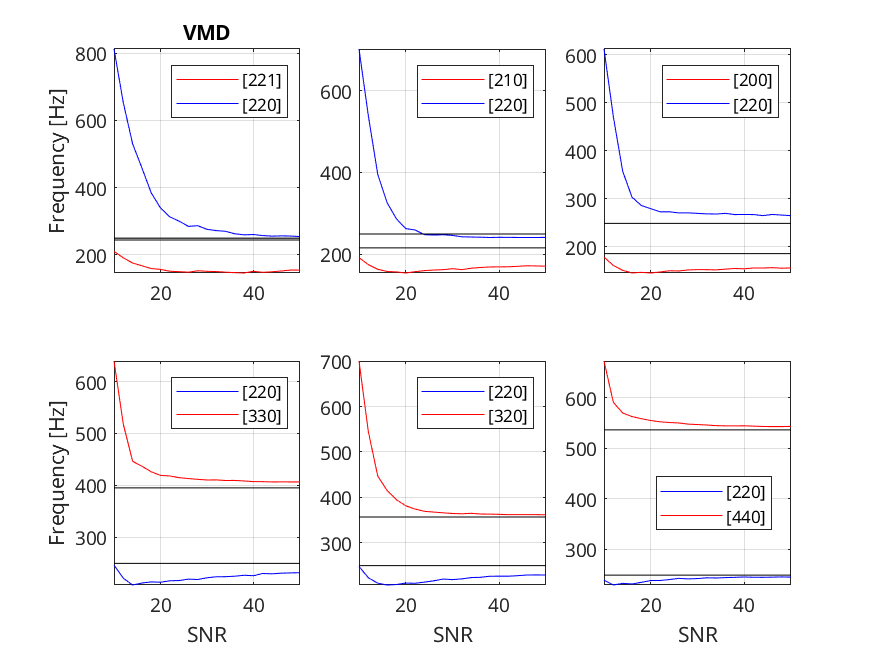}
  \caption{Low SNR regime.}
\end{subfigure}
\begin{subfigure}{.49\textwidth}
  \centering
  \includegraphics[width=\linewidth]{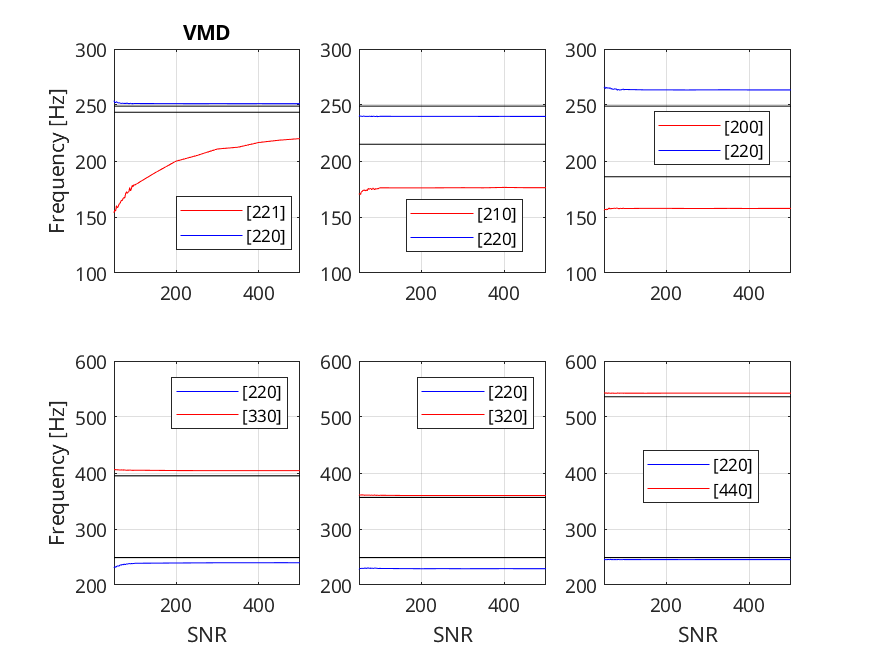}
  \caption{High SNR regime.}
\end{subfigure}
\caption{Application of variational mode decomposition (VMD) to resolve Kerr black hole quasinormal modes in noise.}
\label{fig:VMD_sim}
\end{figure}

Figures~\ref{fig:EMD_sim} and~\ref{fig:VMD_sim} summarize the performance of the EMD and VMD methods, respectively. Both methods successfully resolve the two modes in all the cases considered; however, VMD consistently yields more accurate and tightly bounded frequency estimates.

We divide our analysis into two distinct SNR windows. The low-SNR regime ($\mathrm{SNR} \leq 50$) applies to current second-generation ground-based interferometers. In contrast, the high-SNR regime ($\mathrm{SNR} \leq 500$) targets future third-generation detectors like the Einstein Telescope~\citep{Punturo2010ET} and Cosmic Explorer~\citep{Reitze2019CE}, as well as space-borne missions including LISA~\citep{AmaroSeoane2017LISA}, Taiji~\citep{Taiji}, and TianQin~\citep{TianQin}.

As demonstrated in Figure~\ref{fig:VMD_sim}, the fundamental mode frequency is recovered with high accuracy for $\mathrm{SNR} > 20$. Highly accurate mode frequency estimation is reliably achieved within the high-SNR framework for modes $[3\ 3\ 0]$, $[3\ 2\ 0]$, and $[4\ 4\ 0]$. In contrast, the frequency of the $[2\ 2\ 1]$ overtone remains poorly constrained due to its close spectral proximity to the dominant fundamental frequency. In most unresolved scenarios, the higher-frequency modes show a systematic upward bias, whereas the lower-frequency modes are typically underestimated.

Finally, we evaluate the performance of the proposed IFAD technique introduced in Section~\ref{ssec:IFi}. To estimate the mode frequency using this approach, we estimate the distribution of discrete periods output by the IFAD algorithm and define the mode frequency as the inverse of the period value corresponding to the maximum peak of the distribution.

\begin{figure}[ht]
\centering
\begin{subfigure}{.49\textwidth}
  \centering
  \includegraphics[width=\linewidth]{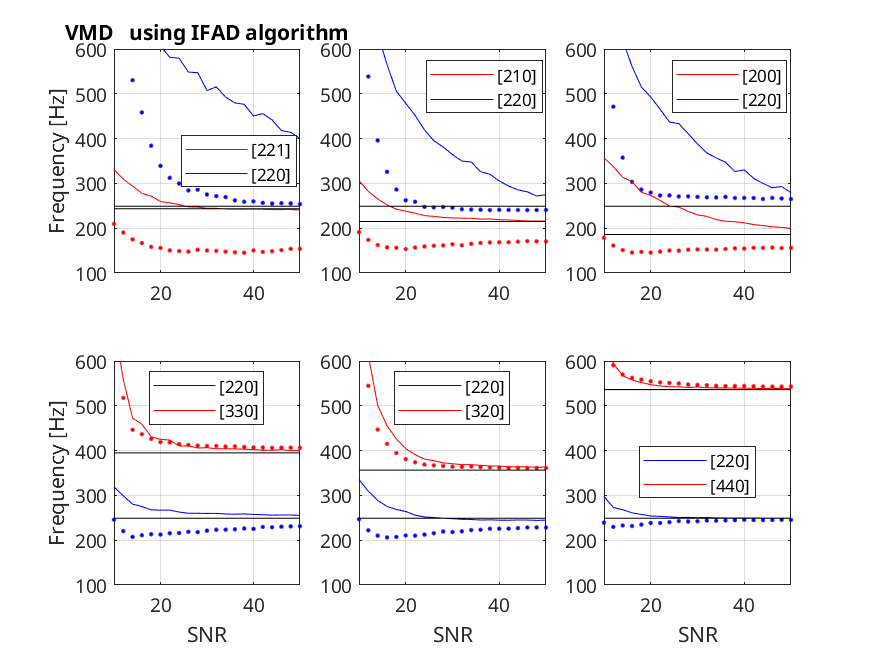}
  \caption{Low SNR regime.}
\end{subfigure}
\begin{subfigure}{.49\textwidth}
  \centering
  \includegraphics[width=\linewidth]{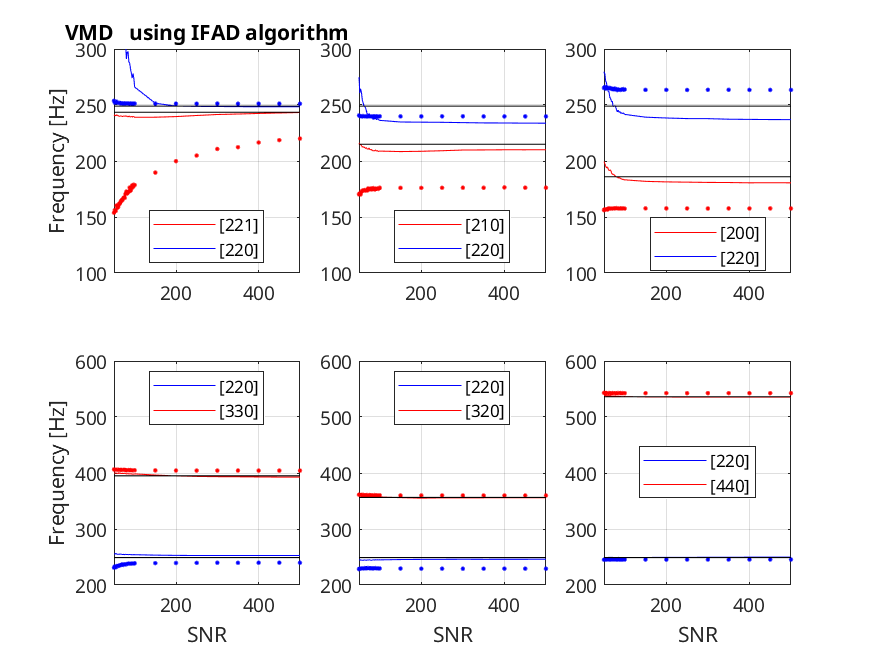}
  \caption{High SNR regime.}
\end{subfigure}
\caption{Application of variational mode decomposition (VMD) to resolve Kerr black hole quasinormal modes in noise, with instantaneous frequencies estimated via the \texttt{IFAD} algorithm. The dotted lines show the baseline estimation tracks from Figure~\ref{fig:VMD_sim} obtained using the standard Hilbert transform.}
\label{fig:VMD_Max_sim}
\end{figure}

The compiled results are shown in Figure~\ref{fig:VMD_Max_sim}. We observe a notable improvement in frequency accuracy compared to the standard Hilbert transform across almost all configurations in both SNR regimes. However, performance degradation is visible at lower SNRs, primarily driven by the small number of distinct oscillation cycles available before the quasinormal mode signals decay into the noise background.

\section{Conclusions}
\label{sec:conclude}

We have investigated the use of adaptive signal decomposition methods for the analysis of post-merger gravitational-wave signals, with a particular emphasis on the challenging problem of resolving multiple quasinormal modes. Using Monte Carlo simulations of noisy ringdown signals composed of pairs of modes, we assessed the performance of Empirical Mode Decomposition (EMD) and Variational Mode Decomposition (VMD) in combination with instantaneous frequency estimation via the Hilbert transform and our proposed IFAD technique.

Our results show that both decomposition methods are capable of separating two-mode signals across a broad range of signal-to-noise ratios. However, VMD consistently provides more accurate and stable frequency estimates than EMD. The analysis highlights several key challenges: (i) the limited duration of ringdown signals restricts the number of observable oscillation cycles, (ii) closely spaced mode frequencies hinder reliable mode separation, and (iii) boundary effects significantly affect instantaneous frequency estimation at early and late times.

To mitigate these issues, we introduced a modified instantaneous frequency estimator, IFAD, which improves accuracy in the high-SNR regime relevant for next-generation gravitational-wave detectors. In this regime, accurate recovery of both the fundamental and secondary mode frequencies becomes feasible for a range of mode combinations. In contrast, performance degrades for lower SNRs and for modes with frequencies close to that of the dominant mode.

These findings have direct implications for black-hole spectroscopy. While current-generation detectors may allow only limited multi-mode measurements, future detectors with higher sensitivity are expected to enable robust detection and characterization of multiple quasinormal modes. The methods investigated here provide a complementary, model-independent approach to ringdown analysis and may play an important role in extracting physical information from high-SNR gravitational-wave observations. Future work will focus on extending these techniques to more realistic signals, including multiple simultaneously excited modes, non-Gaussian noise, and the incorporation of prior physical information to further improve mode identification and parameter estimation.

\section*{Acknowledgments}
This research was funded by the Polish National Science Center Grant No.\ 2023/49/B/ST9/02777. M.T.\ thanks the National Institute for Space Research (INPE, Brazil) for their kind hospitality while this work was done. Conflicts of Interest: The authors declare that they have no competing interests.

\section*{Data access statement}
No data were created or analysed in this study.

\bibliographystyle{apsrev}
\bibliography{refs}
\end{document}